%% file: main.tex
\documentclass[acmtog]{acmart}
\acmSubmissionID{612}

\usepackage{booktabs} 
\usepackage{rotating}
\usepackage{wrapfig,graphicx}
\usepackage[caption=false]{subfig}
\usepackage{soul}
\usepackage{ulem}
\usepackage{svg}
\usepackage{xspace}

\acmPrice{15.00}

\setcopyright{acmcopyright}\acmJournal{TOG}
\acmYear{2022}\acmVolume{41}\acmNumber{4}\acmArticle{149}\acmMonth{7} \acmDOI{10.1145/3528223.3530162}

\setcitestyle{square}

\input{macros.tex}
\begin{document}

\title{Computational Design of Passive Grippers}

\author{Milin Kodnongbua}
\email{milink@cs.washington.edu}

\author{Ian Good}
\email{iangood@uw.edu}

\author{Yu Lou}
\email{louyu27@uw.edu}

\author{Jeffrey Lipton}
\email{jilipton@uw.edu}

\author{Adriana Schulz}
\email{adriana@cs.washington.edu}
\affiliation{%
  \institution{University of Washington}
  \city{Seattle}
  \country{USA}}

\renewcommand{\shortauthors}{Kodnongbua et al.}

\input{sec/0_abstract}



\begin{teaserfigure}
  \centering
  \includegraphics[width=\textwidth]{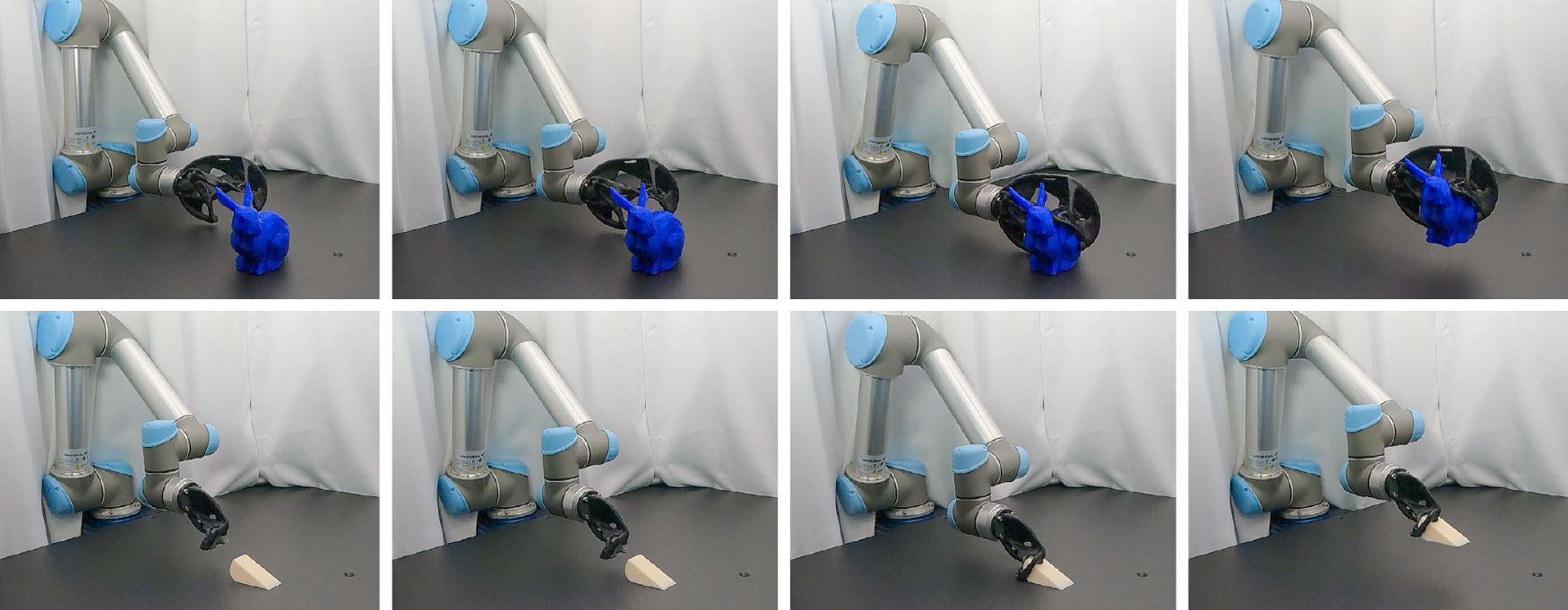}
  \caption{We present an automated algorithm for designing passive grippers given a target object and its positioning. As our algorithm co-designs both the gripper shape and the insert trajectory, our approach broadens the space of shapes that can be passively grasped, compared to existing methods. The figure shows two of the 21 grippers (out of 23) that can successfully pick up the object in reality.  }
  \label{fig:teaser}
  \vspace{8pt}
\end{teaserfigure}

\maketitle
\input{sec/1_intro2.tex}
\input{sec/2_background.tex}
\input{sec/3_overview.tex}
\input{sec/4_grasp_candidate_generation.tex}
\input{sec/5_optimizations.tex}
\input{sec/6_topology_optimization.tex}
\input{sec/8_results.tex}
\input{sec/10_conclusion.tex}

\begin{acks}
This work was funded by NSF grants 2035717, 1954028, 3922035717, and 2017927; the Office of Naval Research grant DB2240; and the Murdock Foundation Materials Foundry grant 201913596.
Special thanks to Tyler Freitas for his assistance in data collection. 
\end{acks}

\bibliographystyle{ACM-Reference-Format}
\bibliography{passiveGripper}

\end{document}

%% file: macros.tex
\theoremstyle{definition}

\newcommand{\unreachable}{{unreachable}\xspace}

\newcommand{\FFO}{FFO\xspace}

\renewcommand{\vec}[1]{\mathbf{\boldsymbol{ #1 }}} 

\newcommand{\stkout}[1]{\ifmmode\text{\sout{\ensuremath{#1}}}\else\sout{#1}\fi}

\newif\ifseechanges

\ifseechanges
\newcommand{\revision}[2]{\textcolor{red}{\stkout{#1}}\textcolor{blue}{#2}}
\else
\newcommand{\revision}[2]{#2}
\fi

%% file: sec/0_abstract.tex
\begin{abstract}

This work proposes a novel generative design tool for passive grippers---robot end effectors that have no additional actuation and instead leverage the existing degrees of freedom in a robotic arm to perform grasping tasks. Passive grippers are used because they offer interesting trade-offs between cost and capabilities. However, existing designs are limited in the types of shapes that can be grasped. This work proposes to use rapid-manufacturing and design optimization to expand the space of shapes that can be passively grasped. 
Our novel generative design algorithm takes in an object and its positioning with respect to a robotic arm and  generates a 3D printable passive gripper that can stably pick the object up. To achieve this, we address the key challenge of jointly optimizing the shape and the insert trajectory to ensure a passively stable grasp. We evaluate our method on a testing suite of 22 objects (23 experiments), all of which were evaluated with physical experiments to bridge the virtual-to-real gap. Code and data are at \url{https://homes.cs.washington.edu/~milink/passive-gripper/}

\end{abstract}

%% file: sec/1_intro2.tex
\section{introduction}

Passive grippers are end-effectors with no actuation. They leverage the existing degrees of freedom in a robotic arm to perform grasping tasks. Among the robotics literature, there has been a growing interest in such grasping techniques as they:  reduce the manufacturing and deployment cost when compared to active grippers; have lower energy consumption since they do not require continuous power; and enable easier human-robot-interaction as anything passively grasped can be removed by a human without changing the robot state. A fundamental limitation to the widespread use of passive grippers, however, is that they are currently significantly restricted in the types of shapes that can be grasped.

Motivating this work is the insight that rapid-manufacturing and design optimization can be used to expand the space of shapes that can be passively grasped. We propose an algorithm that, given an input shape, can automatically generate a 3D printable custom gripper and an accompanying grasp insertion trajectory for stable pickup. This approach is directly applicable to a plethora of robotics applications that target repetitive tasks. For example, it can be used for task-specific tooling of robotic work cells in assembly lines, allowing assembly lines instrumented with simple robotic systems (i.e. systems that lack dexterity) to be easily updated to respond to changes in the product. We can simply optimize and print new passive grippers overnight and re-purpose existing hardware.

The challenge with optimizing a passive gripper for a given input is that (1) there is a valid insert trajectory for making contact with the object and (2) once contact is made, the object can be lifted stably. This implies a large-scale search, nominally at the resolution of the 3D printer, where performance evaluation is expensive to compute since it involves physical evaluation and a nested optimization.  

We address these challenges with two fundamental insights that allow us to reduce the complexity of this search. First, we observe that stability is a function of the contact points between the gripper and the object. We call this set of contact points the \emph{grasp configuration} (GC) and we expose it as a design variable that should be optimized. We observe that, given the GC, an insert trajectory must ensure that there exists a gripper whose final insert configuration touches the GC and can be inserted without colliding with the rest of the shape. Based on this observation, our second insight is that we can search for such a trajectory by creating an abstraction of the gripper geometry that reflects such minimal requirements and then running a co-optimization over the space of trajectories and abstracted gripper shapes. The abstraction we propose is a parametric skeleton of infinitesimal thickness that connects the points on the GC to the \emph{flange frame's origin} (\FFO) of the robot, the center of where we mount the gripper. This reflects the requirement of a rigid object that is attached to the \FFO and makes contact with the selected points.

From these two insights, one could argue for a two-step approach that optimizes the GC for stability and then, co-optimizes the trajectory and the gripper abstraction given the GC. The challenge, however, is that the chance of successfully finding a feasible insertion trajectory depends on the choice of the GC. We therefore propose a strategy for computing a ranked list of stable GC candidates that are likely to enable a collision-free insert trajectory. We use this ranked list of candidates for co-optimizing the trajectory and gripper abstraction. Once the trajectory and the GC are found, a straightforward modification of classic topology optimization algorithms can be used for computing the gripper geometry.

We evaluate our method over an experimental dataset of 23 examples, containing both representative samples from standard grasping datasets as well as a set of challenge models that are difficult to pick up. Our method finds solutions for all of these examples (see Figure~\ref{fig:teaser}). We evaluate the virtual-to-real gap by running real physical experiments in \emph{all} 23 results. We are able to pick up all but two models whose 3D representation fundamentally differed from the real shape. The vast majority of models have high grasp reliability and a large range of stability, but for one we generated a solution with marginal stability.

%% file: sec/2_background.tex
\section{Related Work}
\label{sec:related-work}

\paragraph{Robotic Co-design}
Robot design involves specifying both geometry and actuation sequences. Traditional methods start with general-purpose geometry and then customize the actuation. This is useful given the cost of manufacturing and the relative ease of re-programming. However, the revolution in digital fabrication and the resulting ease of customization has opened a new era of task-specific robot design. A body of recent work has shown the advantages of jointly optimizing a robot's shape and actuation for a variety of tasks such as ground locomotion~\cite{ha2017joint,digumarti2014concurrent,zhao2020robogrammar,luck2020data,spielberg2019learning}, flying~\cite{du2016computational}, swimming~\cite{Pingchuan:diffAqua}, and grasping~\cite{deimel2017automated,xu2021end,pan2020emergent,hazard2018automated,chen2020hardware}. 
Our work builds on this new trend, but instead of customizing a whole robot, we propose to enhance a general-purpose robot with customized end-effectors that can be rapidly fabricated, lowering the cost of customization.

\paragraph{Generative Gripper Design}
Past work on generative gripper design has focused on active grippers. Antipodal grasping was an early target for computational design efforts~\cite{brown19993,velasco1998computer}, with researchers developing shaped fingertips for antipodal grasping using direct formulaic approaches~\cite{brown19993,velasco1998computer,honarpardaz2017finger,schroeffer2019automated,honarpardaz2017fast} and neural network techniques~\cite{ha2020fit2form}. Vacuum-based gripper \revision{}{design} is another direction. The user specifies target locations, and then a 3D printable manifold and superstructure are generated for a specific object~\cite{anubis,fabaloo}. Our work extends these ideas to passive grippers.

\paragraph{Passive Grippers}
In literature, many grippers were classified as passive only because they passively conform to an object (as in the case of compliant~\cite{crooks2017passive} and soft robotics~\cite{seibel2020gecko}). Some so-called passive grippers rely on actuators to release a grasp~\cite{seibel2020gecko,petkovic2013development,nagaoka2018passive,zhang2020compliant}. Truly passive grippers have zero degrees of freedom and have no reliance on actuators. \revision{}{The most widely deployed passive grippers in industry are forklifts, but they can only be used on objects specifically designed for them, e.g., pallets, shipping containers, and FIBC bags. The passive grippers proposed by~\citet{mucchiani2018object,mucchiani2021dynamic} use a rotational motion to engage an antipodal grasp, but they can only pick up extruded cross sections (e.g., a cylinder), the key category not covered by our method. 
Other work has focused on passive deformations around objects to apply antipodal forces~\cite{brodbeck2015extendible,gupta2019design}.}

\begin{table}[t]
\vspace{4pt}
\caption{Comparisons of different grippers on types of objects they are able to pick up.}
\label{tab:SOTA}
\vspace{-4pt}
\resizebox{\linewidth}{!}{%
\begin{tabular}{@{}rcccc@{}}
\toprule
                           & Internal      & Antipodal       & Grips         & Extruded  \\
                           & Pickup & Resistant & Outside CoM & Cross Section  \\ \midrule
Ours                  &  \checkmark             & \checkmark  & \checkmark      &  \\
Non-Actuated Fork Lift     & \checkmark              &             &      &  \\
Mucchiani et al. 2018/2021 &     &  &  \checkmark               &  \checkmark          \\
Antipodal (active or passive)     &   &             &  \checkmark  &  \checkmark          \\
Antipodal (active) + Custom fingertips     &   & \checkmark            &  \checkmark  &  \checkmark          \\
Vacuum-based (active)    &   & \checkmark            &  \checkmark  &  \checkmark   \\
\bottomrule
\end{tabular}%
}
\end{table}

\begin{figure*}[h!]
    \centering
    \includegraphics[width=\linewidth]{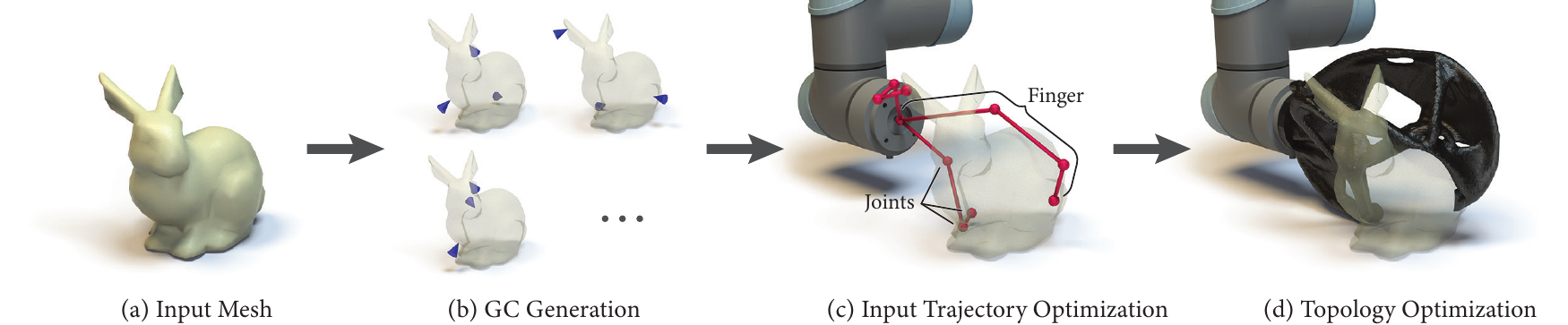}
    \caption{Steps of our algorithm: (a) Import the object's shape from a user provided file; (b) Generate multiple promising GC candidates (where the blue cones point to contact points); (c) Optimize gripper shape and trajectory by using a skeleton (shown in red) as the simplified gripper model; (d) Generate the final gripper shape (shown in black) using topology optimization.}
    \label{fig:overview}
\end{figure*}

Table~\ref{tab:SOTA} summarizes the object shape restrictions posed by different types of grippers. The forklifts can pick up objects around their center of mass (CoM) and through internal pickup (e.g., holes and handles). Active parallel jaw grippers or other passive grippers that generate antipodal grasps work well on a variety of objects because they can create opposing forces anywhere and do not rely on gravity at the CoM for the opposing force. However, they suffer from picking up objects via internal structures and objects that are antipodal resistant. For example, the bottom row of Fig.~\ref{fig:teaser} shows an object that is graspable outside its CoM and is antipodal resistant. Parallel jaw grippers with custom fingertips may be able to pick up the antipodal resistant part due to their large conformable gripping area. Vacuum-based gripper can pick up a wide variety of objects with large enough flat surface. We stress that Table~\ref{tab:SOTA} highlights \emph{characteristics} of objects, not a strict classification. By covering a larger number of characteristics, our work significantly expand the space of objects that can be passively grasped.

\paragraph{Grasping}
All grippers require a grasp planning step and numerous methods have been applied to solving the problem. A fundamental metric for evaluating a grasp is force closure. A grasp is force closed if it can resist any arbitrary set of forces and torques ~\cite{bruyninckx1998generalized,bicchi2000robotic}. A grasp can be partially force closed if it resists all but a subset of forces and torques~\cite{kruger2011partial}. An alternative is to evaluate the caging of an object kinematically, in which an array of obstacles is placed around an object to limit the range of motion of the object~\cite{rodriguez2012caging,seo2016theory}. The downside is that one cannot be sure where the object is relative to the robot when grasped, therefore we chose to focus on a grasp being force closed.

The challenge with physics-based methods for grasp point selection is the need to model noise. Deep learning techniques have been shown to generate robust grasp poses both in the constrained 2.5D bin picking setting~\cite{Mah17Dex} and even the full cluttered 6D grasp pose setting~\cite{Mousavian2019,Mur206D,ten2017grasp,maitin2010cloth,seita2019deep,qin2019keto}. An alternative is to learn visomotor policies directly from raw visual data~\cite{Rah18Vis,Lev16End,Nag19Dee}. While learning methods successfully generate grasp candidates, they are trained primarily for antipodal and vacuum grippers and cannot be directly applied to passive grippers.

%% file: sec/3_overview.tex
\section{Overview}\label{sec:overview}

Our algorithm takes an object's geometry and \revision{orientation}{positioning}, along with a robot's kinematic structure as input, and generates (1) a passive gripper geometry that can be 3D printed and (2) a collision-free insert trajectory for creating a stable grasp.

Our algorithm involves three steps, as illustrated in Figure~\ref{fig:overview}.First, we generate a ranked list of GC (grasp configuration) candidates. Next, we search for a collision-free trajectory for a given GC. The trajectory search is done by jointly optimizing the trajectory and a gripper abstraction. Finally, given the GC and the insert trajectory, we optimize the gripper shape to minimize compliance and weight using topology optimization.

As discussed, GC selection is uniquely responsible for final stability. However, it also impacts the likelihood for finding a valid insert trajectory in the second step and the possible trade-offs between compliance and weight in the third step. We consider four metrics: (1) a binary metric that validates static stability under gravity; (2) a measurement of robustness--the minimum disturbance force and torque that make the object unstable; (3) a heuristic to identify GCs that are unlikely to have a valid insert trajectory; and (4) an estimate of the final gripper weight. We use (1) and (3) to prune randomly sampled GCs and use (2) and (4) to rank the remaining GCs.

The next step is to optimize the insert trajectory to avoid collisions with the object. We consider one GC from the ranked list at a time and proceed to the next best GC if we fail to find the trajectory. Collision avoidance depends on both the trajectory and the gripper's overall shape, making it the most challenging objective. We first simplify the problem by ignoring the structural cost. We observe that the geometric aspect that affects collision is how the contact points are connected to the \FFO and how the connecting structure moves through time. We propose to represent the gripper geometry using a \textit{skeleton} (See Fig. \ref{fig:overview}c.), which is parameterized curves that connect the FFO to each grasp point. We then jointly optimize over the space of skeletons and trajectories by minimizing a cost function associated with collision. We also add a trajectory complexity regularization term to the cost function to reward a simpler trajectory which reduces the chance of collision, leaves more collision-free space to optimize gripper shape, and lowers the robot's energy consumption.

The final step is to compute the gripper geometry. While the skeleton could serve as a workable gripper, it is too fragile. An alternative is to compute the volumetric region that does not intersect with the target object during insert motion, and use this full \emph{collision-free volume} as the gripper. However, this gripper is too heavy. We therefore propose to use topology optimization to generate the final gripper geometry, simultaneously minimizing compliance and weight, while constraining the gripper to lie in the collision-free volume (See Fig. \ref{fig:overview}d.). 

%% file: sec/4_grasp_candidate_generation.tex
\section{GC Candidates Generation}

Recall that a GC (grasp configuration) is a set of three contact points on the target object. We choose three because it is the minimum number of points that constrains a rigid body. More contact points will kinematically over-constrain the solution, make it harder to find a collision-free trajectory, and make the gripper more sensitive to the approach direction \cite{10.1115/1.800857}. Our proposed strategy  for generating this list of candidates involves three steps. First, we propose a method for sampling GCs. Our sampling scheme identifies points on the surface that do not make contact with the floor and that can be directly connected to the \FFO without colliding with the object; sets of three identified points are selected uniformly at random. We then drop GCs that are not statically stable under gravity (Section~\ref{sec:stability}) and the ones that we identify as \emph{\unreachable}---i.e., candidates that are unlikely to have a collision-free gripper shape and grasp trajectory (Section~\ref{sec:reachability}). Finally, we rank the remaining GCs to maximize the robustness to external disturbances once the object is grasped, while trading-off compliance and weight of the gripper shape by using another heuristic metric (Section~\ref{sec:ranking}).

\subsection{Static Stability Under Gravity} \label{partial-minimum-wrench}
\label{sec:stability}

Each contact point makes a contribution to the object's stability by exerting forces onto the object at different angles and location. We assume Coulomb's model of friction that limits the angle with respect to the surface at which the forces can be exerted. For contact point $i$, $| \mathbf{f}_i^T | \le \mu_i | \mathbf{f}_i^N | $ where $\mathbf{f}_i^T$ and $\mathbf{f}_i^N$ are the tangential and normal forces exerted, respectively; and $\mu_i$ is the coefficient of friction at that point. This constraint can be viewed as a cone and can be approximated using a polyhedral cone with $q$ sides where the allowed force is the non-negative linear combination of the edges of the polyhedral cone, called basis. Let $\mathbf{w}_{ij}$ be the wrench (force and torque) generated by $j$-th basis of the cone at point $i$. The contribution from the $i$-th contact point can be written as $\sum_{j=1}^q k_{ij}\mathbf{w}_{ij}$ where $k_{ij} \ge \mathbf{0}$.

We determine if the GC is stable using \textit{partial force closure}, which specifies whether a grasp can withstand a particular external disturbance (in our case, gravity) \cite{partial-force-closure}.
Formally, the partial force checks if there exists $k_{ij} \ge 0$ such that $\sum_i^3\sum_j^q k_{ij}\mathbf{w}_{ij} = [\mathbf{g} \; \mathbf{0}]^T$ where $\mathbf{g}$ is the unit vector of gravity. We \revision{therefore}{} say a GC is \emph{stable} if it meets the partial force closure condition.

There is a slight problem with using this formulation of force contribution for passive grasping because the force of gravity does not directly cause top-facing contact points to generate forces, instead, the gripper only generates a contact force if there is a torque around the center of mass. To account for this, we set the frictional force generated by the top-facing contact points to zero and set those generated by the bottom-facing points depending on the angle with respect to the ground. Formally, we set $\mu_i = \max(0, \mathbf{n}_i \cdot \mathbf{g})\mu$ where $\mu$ is the base coefficient of friction.

There is a stronger condition called \textit{force closure} which specifies whether a grasp can withstand any external forces and torques~\cite{bruyninckx1998generalized,bicchi2000robotic}.
However, this type of grasp which typically involves contact points on different sides of the object makes breaking contact more difficult.
We observe that, by utilizing gravity, the object can still be held stable without a force closure grasp.
Hence, the partial force closure condition is sufficient for our purpose.

\begin{figure}[t]
    \centering
    \includegraphics[width=0.85\columnwidth]{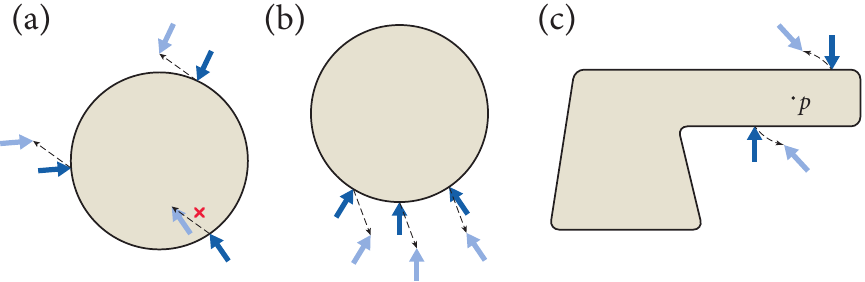}
    \caption{Illustrations of the reachability heuristic. The dark blue arrows point to the contact points. The dashed black arrows show a candidate rigid motion to break all contact points simultaneously. (a) contact points cannot instantaneously break contact with the circle; (b) contact points can break the contact by translation; (c) contact points can leave the object by rotating counterclockwise around the point $p$.}
    \label{fig:contact-point}
\end{figure}

\subsection{Identifying Unreachable GC}
\label{sec:reachability}
Given a GC, it is expensive to determine if a collision-free trajectory exists---it involves solving the co-optimization problem in Section~\ref{sec:optimization}. Thus, we need a fast heuristic that identifies \unreachable GCs. Our insight is to check if there exists an instantaneous motion where all contact points simultaneously break contact with the object.

We will motivate our heuristic in 2D for ease of illustration; these ideas extend naturally to 3D. Imagine grabbing a circle at three equally spaced locations in a 2D space (see Fig.~\ref{fig:contact-point}a.). It is trivial to see that there is no rigid motion to break all three contacts simultaneously without colliding with the circle, so we can drop it without running the expensive optimization. However, if the contact points are roughly on the same side (see Fig.~\ref{fig:contact-point}b.), then we can break the contacts just by moving the contact points away; hence, we keep this GC.
One might be tempted to just check if the normals of each contact point are pointing in the same hemisphere. This is too strict. Consider grabbing the object in Fig.~\ref{fig:contact-point}c. The two contacts' normals are pointing in the opposite direction. However, this GC is reachable since we can rotate counterclockwise around the midpoint between the two contact points.

Our heuristic searches for an instantaneous rigid motion (translation and rotation) so that all contact points simultaneously sufficiently break contact with the object. 
We define the instantaneous motion of the gripper as $(\vec{v}, \vec{\omega}, \vec{c})$, where $\vec{v}$ is the instantaneous translational velocity and $\vec{\omega}$ is the instantaneous angular velocity around the rotation center $\vec{c}$.
We can derive the instantaneous velocity of a single contact point as $\vec{v}_i = \vec{v} + \vec{\omega} \times (\vec{c}_i - \vec{c})$ where $\vec{c}_i$ is position of the contact point. The contact point breaks contact if $\vec{v}_i \cdot \vec{n}_i \ge 0$ where $\vec{n}_i$ is the normal direction at the contact point. We also limit the angle between $\vec{v}_i$ and $\vec{n}_i$ to $\theta_\text{max}$ to avoid nearly parallel motion to the surface. Formally, if there does not exists a motion $(\vec{v}, \vec{\omega}, \vec{c})$ such that $\vec{v}_i \cdot \vec{n}_i \ge \lVert \vec{v}_i \rVert \cos(\theta_\text{max})$ for all $i$, then the GC is \emph{unreachable}.

Inspired by the barrier method~\cite{barrier-method}, we construct a loss function that penalizes motions that violate the constraints\revision{ and solve}{. We then solve the following minimization problem} for a valid motion \revision{with}{using} a standard gradient-based optimization:
\begin{align*}
    \min_{(\vec{v},\vec{\omega},\vec{c})} \sum_i \big[ [\cos(\theta_\text{max}) - \vec{v}_i \cdot \vec{n}_i]_+ + [\lVert \vec{v}_i \rVert - 1]_+ \big]
\end{align*}
where $[\cdot]_+$ denotes $\max(0, \cdot)$.

Note that we do not normalize $\vec{v}_i$ to avoid numerical instability when the value approaches zero. The angle between two vectors will not exceed $\theta_\text{max}$ because we also constrain $\lVert \vec{v}_i \rVert \le 1$. A GC is said to be \textit{unreachable} if the loss function does not converge to zero.

\subsection{Ranking}
\label{sec:ranking}

The final step is to produce a ranked list of candidates for the next stage. We consider two metrics.

The first metric evaluates the minimum disturbance force that causes the object to become unstable. This metric measures how stable the grasp will be after the object is picked up and experiencing the transfer motion. Inspired by the well-known minimum wrench metric, we introduce a new metric called \textit{partial minimum wrench}, which measures the minimum additional external wrench to violate the partial force closure condition. Intuitively, this metric corresponds to the maximum tilt angle and/or acceleration that an object held by this GC can experience before falling off. A higher partial minimum wrench is better.

The second metric is the estimated finger length. If the contact points are far from the \FFO, the gripper will be cumbersome and the path to reach the point will be long. We therefore estimate the finger length by finding the shortest non-colliding path from the \FFO to each contact point and taking the maximum path length. A shorter length is better.

Since the two metrics can be conflicting, our ranking scheme is inspired by the non-dominated sorting criteria in multi-objective optimization~\cite{deb2000fast}. 
We segment the candidates into multiple Pareto frontiers, and sort them by the estimated finger length within each frontier.

%% file: sec/5_optimizations.tex
\section{Trajectory Optimization}\label{sec:optimization}

\newcommand{\optcfg}{OP}

In this stage, for each GC candidate, we want to search for an insert trajectory that avoids collision. Since collisions depend both on the gripper's shape and the trajectory, we need to design them together. The fundamental challenge in the co-design of a gripper and trajectory is the complexity of the search space. The nominal search space of the gripper geometry corresponds to the resolution of the voxel grid that a 3D printer can afford, and the search space of a trajectory can be described by the degrees of freedom of a robotic arm over time. We propose a novel reduced representation of the search space to make this search tractable.

\subsection{Gripper Abstraction}
We represent a gripper design as a \textit{skeleton} which is comprised of three \textit{fingers}, each with multiple \textit{joints} connecting the \FFO to the three contact points (Fig. \ref{fig:overview}c). This representation stems from the weakest constraint that the grasping points need to be connected to the \FFO. We use a skeleton of infinitesimal thickness to evaluate collisions over 2D curves as opposed to volumes; we account for manufacturability by expanding the mesh by an offset that corresponds to half of the printer resolution.

The insert trajectory is represented as a linear interpolation between a list of robot states in joint space. We call these robot states the \textit{keyframes} of the trajectory. The first keyframe defines the robot when the gripper is outside of the object's proximity, and the last keyframe defines the robot when making contacts with the target object.

We define the gripper skeleton as $\mathcal{G} \in \mathbb{R}^{3 \times m \times 3}$ where $m$ is the number of joints in each finger\revision{s}{} and define the trajectory as $\mathcal{T} \in \mathbb{R}^{n \times d}$ where $n$ is the number of keyframes and $d$ is the robot's degrees of freedom. Our goal is to optimize the tuple of the gripper skeleton and the trajectory: $(\mathcal{G}(\mathbf{x}), \mathcal{T}(\mathbf{x}))$ where $\mathbf{x}\in \mathbb{R}^N$ are the adjustable parameters, namely the intermediate joint positions for every finger, and intermediate trajectory keyframes, for a total of $N := 3 \cdot(m-2) \cdot 3 +d \cdot (n-2)$ parameters. In our implementation, we set $m=4$, $n=4$ and $d=6$, resulting in 30 degrees of freedom in total. We denote $(\mathcal{G}(\mathbf{0}), \mathcal{T}(\mathbf{0}))$ as the initial guess from the initialization method described in Section~\ref{sec:cooptimization}. We specify appropriate ranges $\sigma \in \mathbb{R}^N$ to each of the adjustable parameters (i.e. $-\sigma_i \le \mathbf{x}_i \le \sigma_i$) to ensure the search space is connected and the solution remains valid (e.g. the robot does not self intersect and the skeleton maintains its overall shape). In the following discussion, we omit $\mathbf{x}$ and hence write $\mathcal{G}$ and $\mathcal{T}$ to mean $\mathcal{G}(\mathbf{x})$ and $\mathcal{T}(\mathbf{x})$.

\subsection{Co-Optimization of Trajectory and Gripper Skeleton}
\label{sec:cooptimization}

To co-optimize the trajectory and the gripper skeleton, we propose an objective function with four energy terms: {gripper collision energy} $E_g$, {trajectory collision energy} $E_t$, {robot collision energy} $E_r$, and the trajectory regularizer $L$. The first three terms focus on collision, while the regularizer penalizes complex trajectories. 

\begin{figure}[b]
    \centering
    \includegraphics[width=\columnwidth]{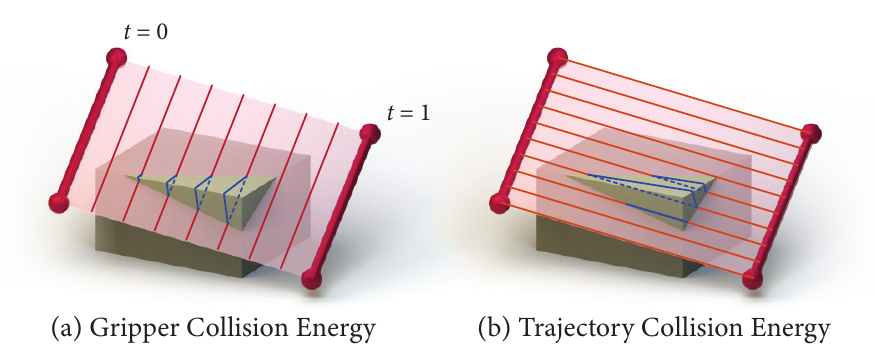}
     \caption{
     Our collision energy measures the collision of the surface swept by the skeleton over trajectory shown in pink. We compute collision using path intersections along two directions: 
     In (a), the path along the skeleton is tested at different time steps in the trajectory (shown by vertical red lines); In (b), the path along the trajectory is tested at different points on the skeleton (shown by horizontal orange lines). Collisions by the paths are shown in blue. Maximum collision of the paths in (a) defines the gripper collision energy, and likewise maximum of those in (b) defines the trajectory collision energy. Note: a simple box and a linear path is used for simplicity of the visualization.}
    \label{fig:traj-gripper-cost}
\end{figure}

The gripper collision energy is defined as the maximum collision at any point of the trajectory measured over the whole gripper skeleton. Similarly, the trajectory collision energy is defined as the maximum collision of a single point in the gripper measured over the whole trajectory. These two metrics are illustrated in Figure~\ref{fig:traj-gripper-cost}. We note that, in theory, either one of these metrics can individually represent the collision error we target. In practice, however, we cannot directly compute them, but must instead evaluate them over a discretization of the trajectory and skeleton. Our insight for combining these metrics is that they describe the two basis directions for the surface, making the evaluation over a discretization robust to small features.

Both metrics depend on a measurement of collision over a path:  the collision of the skeleton at any point can be expressed as the sum of the collision of its fingers (piece-wise linear paths) and the collision of a point throughout the whole trajectory is measured from the path this point traces. We will describe our measurement of collision over a path later in this section.

Since the robot end-effector cannot go lower than a certain clearance height $h$, we introduce a robot collision energy term $E_r$ which is the maximum penetration into the ground by \FFO's position. We penalize the complexity of the trajectory with the L2 norm of the trajectory variations from the initial guess (i.e. the trajectory components in $\mathbf{x}$).

Our optimizer minimizes the following energy term:
\begin{align*}
    \min_{\mathbf{x}} E_g(\mathcal{G}, \mathcal{T}) + E_t(\mathcal{G}, \mathcal{T}) + \lambda_1 E_r(\mathcal{T}) + \lambda_2 L(\mathcal{T})
\end{align*}
where $\lambda_1$ and $\lambda_2$ are the significance of the robot collision energy and the trajectory regularizer. Since this objective is non-differentiable, we use controlled random search (CRS) with local mutation for minimizing this energy~\cite{kaelo2006some}.

\begin{figure}
    \centering
    \includegraphics[width= 0.7\columnwidth]{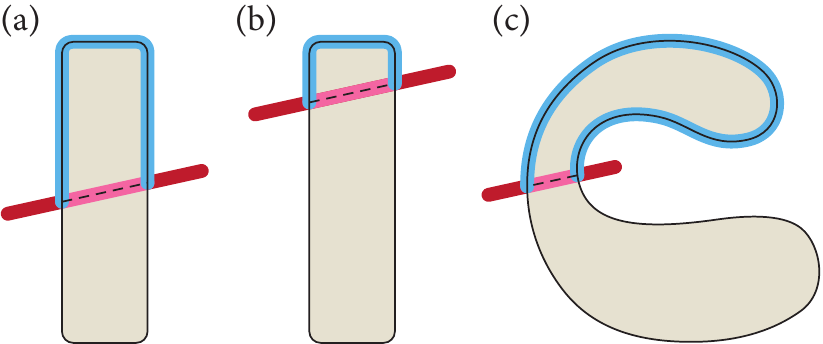}
    \caption{Illustrations of the \textit{inside distance} (pink) and \textit{wrap-around distance} (cyan) of the skeleton (red) through some objects. The path in (b) is closer to a collision-free state than that in (a), but the inside distance is the same. The path in (c) has lower inside distance than (a) and (b), but it needs more work to reach a collision-free state.}
    \label{fig:intuition}
\end{figure}

\begin{figure}
    \centering
    \includegraphics[width=0.9\columnwidth]{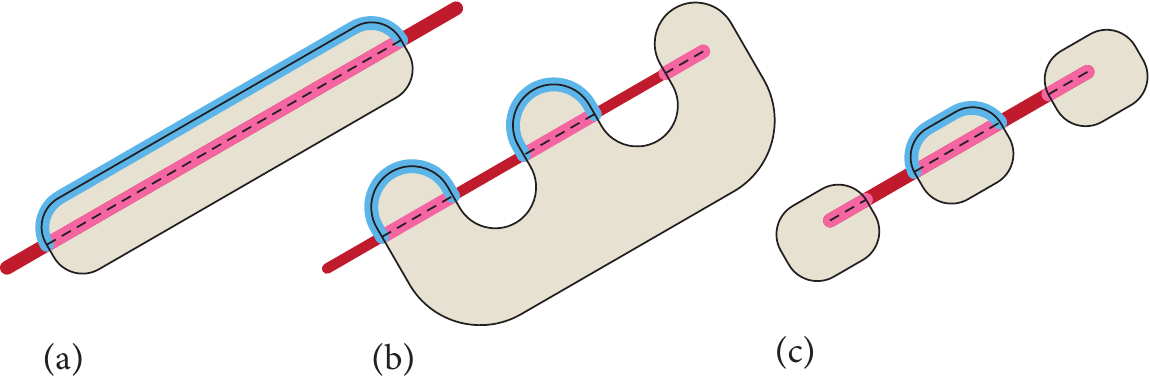}
    \caption{The \textit{wrap-around} distance is the sum of the geodesics (cyan) between every pair of entering and exiting intersections. The path in (a) has one such pair and in (b) has \revision{three}{two}. Some endpoints in \revision{(c)}{(b)} and \revision{(d)}{(c)} do not have a pair and therefore the geodesics are not defined.}
    \label{fig:geodesic}
\end{figure}

\paragraph{Collision Along a Path}
There are several approaches to compute the collision of a given path.
One trivial approach is to use the length of the path that lies inside the object. We call this length the \textit{inside distance}. However, this approach does not capture the degree of collision, as illustrated in Figure~\ref{fig:intuition}. We propose to include a measurement of the shortest additional distance to make the path collision-free. We call it the \textit{wrap-around distance} (See Fig. \ref{fig:intuition}). To measure the wrap-around distance, we compute the geodesic distance between a pair of points when the path enters and exits the object. We take the sum of the geodesics if the path enters and exits multiple times, and we do not include the geodesic if one or both endpoints are in the object. The collision of a path is defined as the sum of the inside and wrap-around distance (See Fig. \ref{fig:geodesic}).

\paragraph{Discretization}
We compute $E_g$ and $E_t$ by sampling the trajectory and skeleton, respectively. We adaptively subdivide the trajectory so that the skeleton moves by at most some distance $d_\text{sub}$ and we adaptively subdivide the fingers by that same distance. This sampling is repeated for every iteration as lengths of individual segments can significantly change when adjustable parameters change.

To find where the path enters and exits the object, we use
a raytracing algorithm to find intersections. Since the path for trajectory is not linear, we adaptively subdivide the path with some linearity tolerance $d_\text{lin}$ into connected line segments.

To efficiently compute the geodesic distance between any given two points, we need to allow some accuracy trade-offs. We approximate the geodesic distance by running a standard shortest path algorithm on the vertices and edges of the triangle mesh. We use the isotropic remeshing algorithm to generate a new mesh whose edge lengths are as close to one another as possible. This makes the approximation more accurate. We precompute the shortest distance between every pair of vertices. The approximate geodesic distance for any given two points on the surface is computed as the distance of those two points to their closest vertices plus the precomputed geodesics between the two vertices.

\paragraph{Initialization}
We initialize the skeleton by computing the shortest, non-colliding curves that connect 
from the \FFO to each contact point and then simplifying these curves to the right number of joints by discretizing the curves, expanding them by the surface normal, and removing vertices that does not contribute to collision avoidance. We initialize the trajectory to be a straight line towards the object. 

%% file: sec/6_topology_optimization.tex
\section{Topology Optimization and Refinement}

\begin{figure}
    \centering
    \includegraphics[width=.5\columnwidth]{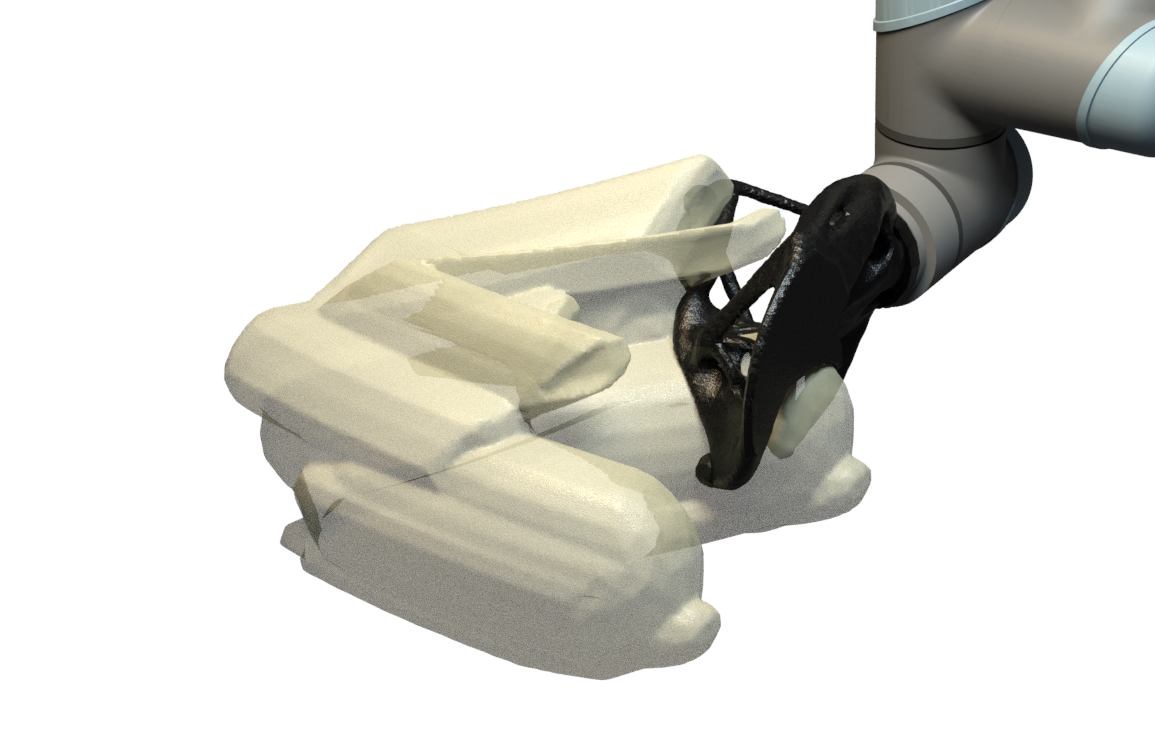}
    \caption{In the robot's reference frame, the swept volume of the bunny (shown in white) is the space occupied by the bunny throughout the trajectory. The complementary space is the \textit{collision-free} space for topology optimization. The topology optimized gripper is shown in black.}
    \label{fig:swept-vol}
\end{figure}

At our last step, we generate the gripper design by performing discrete topology optimization over the \emph{collision-free} volume~\cite{bendsoe2003topology}. Since the trajectory is known, this volume can be computed as the complementary space of the swept volume of an object moving away from the gripper (See Fig. \ref{fig:swept-vol}).
For the boundary conditions, we set the external forces to be the forces exerted by the contact points of the known GC along the normal directions and fix parts of the gripper around the \FFO. In post-processing, we apply a smoothing kernel and run marching cubes to retrieve a smooth mesh. We further refine the gripper geometry by adding a small sphere at each contact point and subtracting the result with the swept volume to ensure an accurate geometry around the contact points. This increases the area of contact and improves the robustness of the grasp. Lastly, we add robot-specific mounting structures such as holes and a mounting plate for fast installation.

%% file: sec/8_results.tex
\section{Results}\label{sec:results}

We evaluate our algorithm on its success rate and demonstrate different designs fabricated in real. We refer the reader to the supplementary video for demonstrations of the grippers in motion.

\subsection{Evaluation Set}

\begin{figure}[b]
    \centering
    \includegraphics[width=\columnwidth]{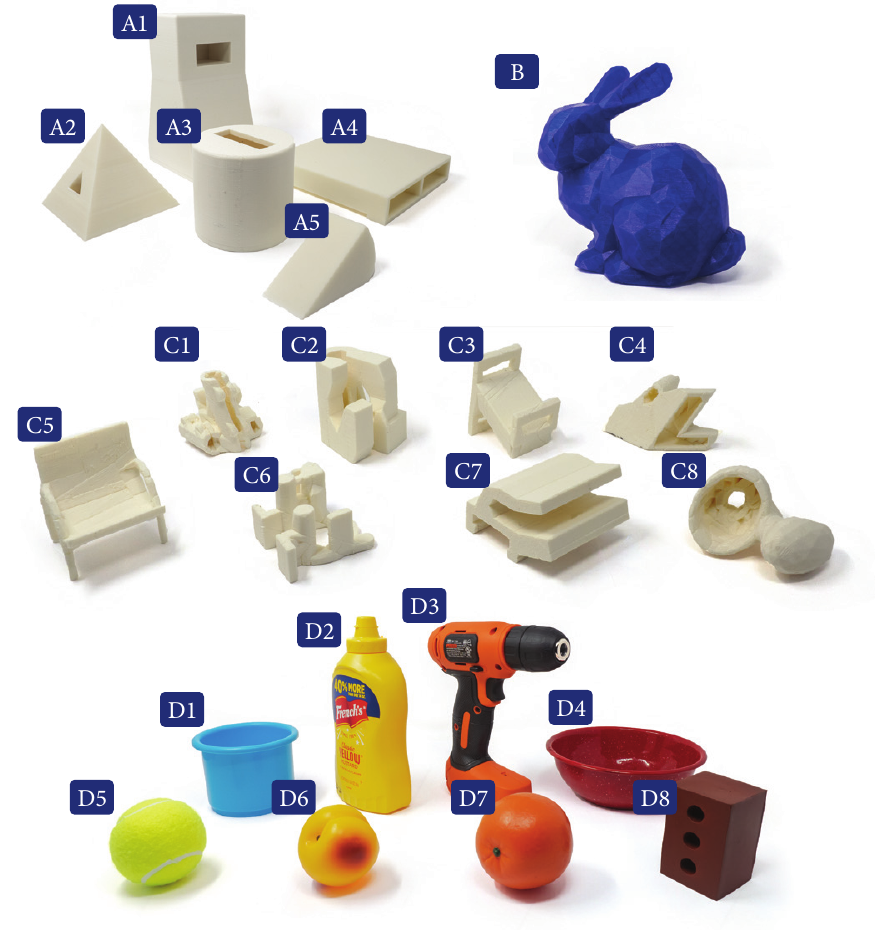}
    \caption{Evaluation set. A: engineered models; B: bunny; C: challenge models; D: samples from YCB dataset.}
    \label{fig:evaluation-set}
\end{figure}

\begin{figure*}[h!]
    \centering
    \includegraphics[width=\textwidth]{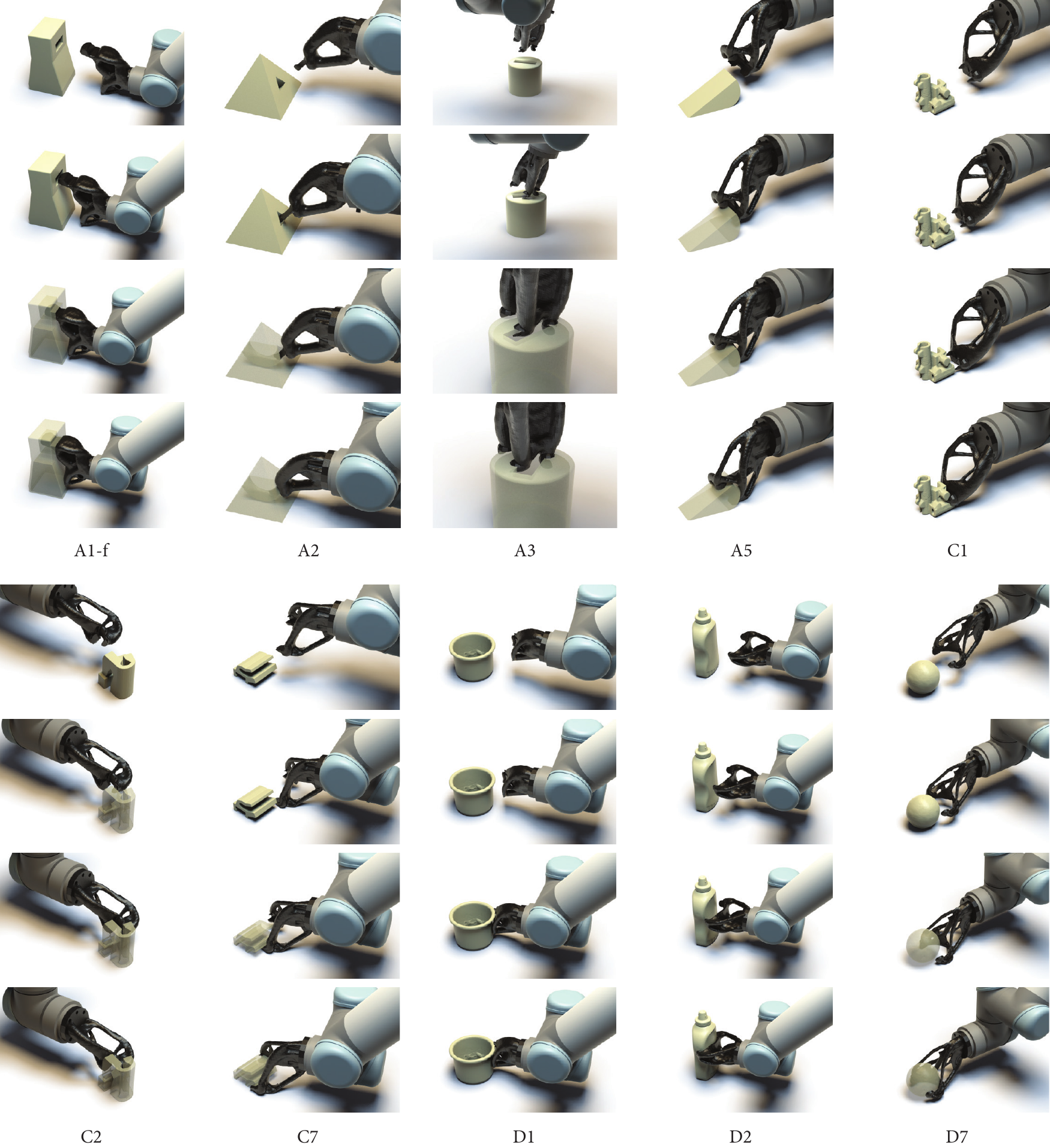}
    \caption{Example grippers and their trajectories from objects in our dataset. Generated grippers and trajectories vary significantly across different models. These gripper designs mimic everyday mechanical tools such as wrenches (A5 and D1) and tongs (C7, D2, and D7), but are highly tailored to fit the objects. The accompanying trajectories can be classified into insert motions (A5, C2, and C7) and twist motions (A2, A3) and are nontrivial to design by hand.}
    \label{fig:trajectory}
\end{figure*}

We created a testing set of 22 objects (See Fig. \ref{fig:evaluation-set}).

Five models are engineered to demonstrate the unique capabilities of the generated passive grippers. A1-A3 are designed to represent internal structures. A1 contains an L-shaped hole at the front. The pyramid (A2) contains a curved hole that requires a twist motion to insert. The \textit{Top Key} (A3) contains a rectangular slot at the top and requires a quarter turn to insert. The pallet (A4) is designed to test if our algorithm can generate a fork-lift. The wedge (A5) is engineered to resist an antipodal grasp because of its tapered shape at all orientations. To the best of our knowledge, objects (A2) and (A5) cannot be passively picked up by any prior work. 

In addition to our engineered models, we include a representative set from prior work.  First, to uphold the SIGGRAPH tradition, we add the Stanford bunny (B) to our evaluation set. We choose some challenging objects (C1-C8) from the  Fit2Form dataset \cite{ha2020fit2form} representing adversarial objects from Dex-Net 2.0 (C1,C2, C4,C6-8) \cite{Mah17Dex} and objects from ShapeNet (C3, C5) \cite{shapenet2015}. We a pick a subset of YCB objects \cite{calli_singh_walsman_srinivasa_abbeel_dollar_2015,calli_walsman_singh_srinivasa_abbeel_dollar_2015} that are feature-rich and contain no moving parts (D1-D8).

The input to our algorithm is the shape and its positioning. We run two experiments with (A1) using orthogonal positions: inserting from the front and from the side, resulting in a total of 23 experiments. We refer to them as \textit{Front Key} (A1-f) and \textit{Side Key} (A1-s).

\subsection{Implementation and Experimental Setup}

Our algorithm is implemented in C++, using libigl \cite{libigl} and CGAL \cite{cgal:eb-21b} for most mesh processing tasks. We use UR5 as our robot, which has six degrees of freedom. We use the code from \citet{swept-volume-siggraph} to calculate swept volume, and we use ToPy \cite{topy} to run topology optimization with the voxel size of 2 millimeters. We ran the algorithm on all 23 experiments on a cluster with 40 cores.

In our GC generation stage, we randomly sample 1,000 points on the mesh and generate 3,000 GCs. We assume the base coefficient of friction $\mu$ is 0.5 in modeling the contacts. For the reachability heuristic, we set $\theta_\text{max}$ to $80^{\circ}$. In computing collision of a path, we set both the subdivision distance threshold ($d_\text{sub}$) and the linearity threshold of the trajectory ($d_\text{lin}$) to 1 millimeter. In computing the total objective function, we set the robot floor clearance ($h$), the robot energy significance ($\lambda_1$), and the regularizer significance ($\lambda_2$) to 0.05 meters, 1000, and 1e-6, respectively. In the optimization stage, we use NLopt's implementation of CRS for trajectory optimization \cite{nlopt} with the population size of 10,000 and relative tolerance of 1e-6. We allow the intermediate finger joints to vary within 1 centimeter, and we allow the six joints to deviate within $5^\circ, 5^\circ, 5^\circ, 45^\circ, 25^\circ$, and $90^\circ$, respectively.

We set the robot grasping position (the last keyframe) so that the end effector is pointing forward and we place the object right in front of the end effector. This is true for all the models except for the Top Key (A3) which we point the end effector downwards.
We ran the optimization stage until we were able to find three candidates that succeeded or until we ran out of grasp candidates, whichever happened first. We picked one candidate to fabricate and validate in the real world.

\subsection{Algorithm Performance}

The GC candidate generation step took less than a minute for each model. The trajectory generation took between 7 and 26 minutes, and 13 minutes on average per GC candidate per model. This step terminated after running on 5 GC candidates on average and succeeded on at least one candidate on every model. The topology optimization took on average 1 hour and 26 minutes, but varied largely with the dimensions of the bounding box.

The whole pipeline takes on average 2 hours and 38 minutes per model, and on average $54\%$ of the time is spent running topology optimization. While several  computation steps could be sped up with straightforward implementation updates (for example, faster topology optimization~\cite{zhu2017twoscale}), our computation speed is on par with SOTA generative design systems and suitable for practical use, particularity when considering manufacturing time.

\subsection{Qualitative Analysis}

As illustrated in Figure~\ref{fig:trajectory}, the gripper shapes and trajectories are highly customized for each particular object. While many trajectories may seem obvious given a gripper, our algorithm can generate non-intuitive solutions. For example, it took one of the authors (who had seen all visualizations) almost two minutes to figure out how one of the grippers fits into its object. This highlights the tightly coupled nature of the problem and the need for co-optimization. 

While all solutions are unique, we note that it is possible to classify gripper shapes into three categories. \textit{Inserts} take most of the load in a single point, typically inside a cavity on the object, as is shown for examples A1-f, A2, A3, C1, and C2. \textit{Tongs} rely on a wide separation of support from underneath, as shown in examples C7, D2, and D7. Finally \textit{wrenches} have a narrow grasp from opposite sides that distributes the load on all three grasp points, as shown in A5 and D1. While customized to particular shapes, the designs replicate standard types of simple tools.

The trajectories most grippers follow also fall into a triadic grouping.
Most objects fall into the \textit{front insert} category where the gripper moves directly in and grabs the object from the bottom. Some examples include the front key (A1), the pallet (A4), the cup (D1), the orange (D7), and the challenge object C1. Although the trajectories may seem simple, they are not trivial to be designed by a user because they can be sensitive to approach angles and sequencing. For example, to get the orange, we need to go slightly down then insert to avoid the convexity of the sphere, and to get the front key (A1), we need the right amount of in and up motion.

A number of designs fall into the \textit{side insert} category where the required motion is to slide horizontally into the desired position. Some examples include our wedge (A5), our Side Key (A1-s), objects C2 and C7. Our algorithm is able to find the solution even if the required motion is not perfectly straight. In the bunny example, the robot needs to slightly turn to get around the convex part. 

Our pyramid (A2) and \textit{Top Key} (A3) fall in to the \textit{twist insert} category. These objects are designed to only be grabbed by rotation motions which our algorithm is able to find. 

Together, these results suggest that our algorithm is successful in understanding the affordances of each object. For example, object A1 is picked up with a front or side motion depending on the orientation of the cavity. Objects with bilateral symmetry are grasped on opposing sides, and objects with recesses have smooth insertion trajectories. 

\subsection{Physical Experimental Setup}

We 3D print the grippers using two materials. We use a material with high coefficient of friction for the contact region (TangoBlackPlus) and a material with high stiffness and low cost for the rest of the griper (ABS). While these could be printed together in a multi-material printer, to reduce cost, we use the Stratasys J750 Digital Anatomy for the contact region and Stratasys FDM 3D Printers (Fortus 250mc, F120, F170) for the rest and assemble the two parts.

\begin{figure} [b]
    \centering
    \includegraphics[width=0.7\columnwidth]{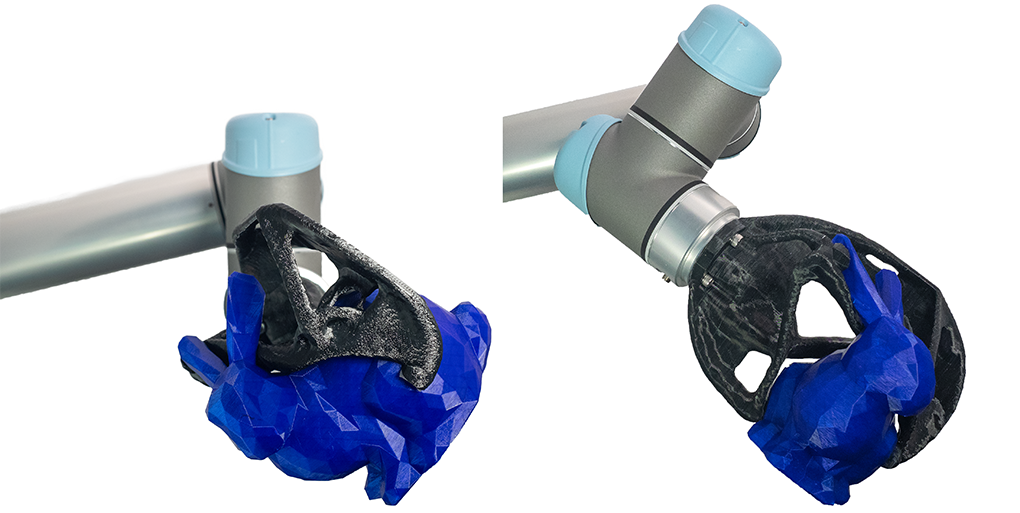}
    \caption{Post-grasp stability test for the bunny. We measure the maximum roll angle both CW and CCW (left) and forward tilt angle (right) before the bunny falls.}
    \label{fig:drop-test}
\end{figure}

All objects and printed grippers were tested on the UR5 arm in moveJ mode with the trajectories specified by our algorithm. We tested object pick up as well as resistance to dropping once the object was correctly seated in the gripper. This involved rotating clockwise, counterclockwise, and a forward roll until the object fully fell out of the gripper and can be seen in Fig.\ref{fig:drop-test}. For the \textit{Top Key} we roll in the opposite direction due to joint limitations in the UR5 at this position. We repeat every test 10 times per object.

\begin{table}[t]
    \centering
    \vspace{4pt}
    \caption{The number of successful pickups and maximum roll and pitch angles before the object falls out of the gripper with ten tests per category per object. For roll and pitch, the object was correctly loaded into the gripper to represent stability once an object was successfully picked up. Some objects (A3, C1, C3, C4, C6, D7) demonstrated multiple falling modes where an object could find additional regions of stability, resulting in large standard deviations.}   
    \label{tab:results}
    \vspace{-4pt}
    \small
\begin{tabular}{lrrrrrrr}
    \toprule
    \multicolumn{1}{l}{Object} &
    \multicolumn{1}{r}{Pickups} &
    \multicolumn{2}{c}{CW (deg)} &
    \multicolumn{2}{c}{CCW (deg)} &
    \multicolumn{2}{c}{Fwd Tilt (deg)}
        \\
    \cmidrule{3-4} \cmidrule{5-6} \cmidrule{7-8}
    & & mean & sd & mean & sd & mean & sd \\
        \midrule
A1-f & 10 & 170 & 3 & -162 & 6 & 68 & 9 \\
A1-s & 10 & 34 & 6 & -23 & 7 & 30 & 4 \\
A2 & 10 & 360 & 0 & -360 & 0 & 53 & 23 \\
A3 & 10 & 360 & 0 & -359 & 3 & -292 & 144 \\
A4 & 10 & 237 & 3 & -28 & 3 & 25 & 2 \\
A5 & 10 & 52 & 6 & -59 & 12 & 34 & 4 \\
B & 10 & 60 & 3 & -194 & 1 & 145 & 3 \\
C1 & 10 & 341 & 34 & -246 & 41 & 118 & 15 \\
C2 & 10 & 38 & 11 & -238 & 5 & 276 & 10 \\
C3 & 10 & 242 & 65 & -189 & 5 & 83 & 3 \\
C4 & 10 & 360 & 0 & -332 & 59 & 316 & 72 \\
C5 & 2 & 149 & 12 & -14 & 4 & 9 & 2 \\
C6 & 6 & 308 & 113 & -358 & 4 & 28 & 69 \\
C7 & 10 & 255 & 22 & -66 & 5 & 34 & 9 \\
D1 & 10 & 62 & 16 & -107 & 44 & 68 & 16 \\
D2 & 10 & 92 & 2 & -101 & 2 & 26 & 5 \\
D3 & 9 & 27 & 2 & -21 & 4 & 19 & 5 \\
D5 & 10 & 127 & 15 & -59 & 3 & 14 & 2 \\
D6 & 10 & 101 & 8 & -89 & 4 & 69 & 33 \\
D7 & 8 & 97 & 6 & -178 & 102 & 40 & 7 \\
D8 & 10 & 50 & 9 & -104 & 10 & 30 & 2 \\
\bottomrule
    \end{tabular}
    \vspace{-10pt}
\end{table}

\subsection{Real World Validation}

Of the 23 experiments, 21 lead to successful pickups and the results are shown in Table~\ref{tab:results}. 17 of these experiments had 100\% success rate and only two had success rates below 80\%. These results, and the following analysis of the failure cases, show that our method is able to successfully bridge the virtual-to-reality gap and displays high grasp reliability in real experiments. Visualization of these results are shown in Figure~\ref{fig:teaser} and our supplemental video.

The two failure cases (D4 and C8, not shown on the table) had significant discrepancies in their virtual representations. Since our gripper and trajectory were computed over the wrong input, the results were not well suited in practice. The 3D mesh we used to represent the bowl (D4) is thinner than the actual bowl, causing the gripper to collide during insertion. The challenge model (C8) was not oriented in its rest position causing inaccuracies during stability analysis. We note that the two models that had 80-90\% success rates had similar discrepancies, but those affected our gripper to a lesser extent because of symmetries in the models and the locations where the discrepancies happened.

The two models with marginal stability (C5 and C6) were examples from the adversarial dataset, which are notoriously challenging to grasp. These failed because the grasp locations we chose did not leave much room for misalignment. The grasp location must be robust to errors in the gripper fabrication (order of 0.5mm), the motion of the robot (order of 0.1mm) and human positioning the object (order of 0.5mm in the best case scenario). This uncertainty was not accounted for by our algorithm that evaluates the stability of a grasp configuration (Section~\ref{sec:stability}). We highlight, however, that our system was robust to these errors in all but these two adversarial examples, which shows the effectiveness of our approach.

\subsection{Limitations and Future Work}

The most evident direction for improvement is stability evaluation for selecting the grasp configurations. This would make our method more robust to errors in object placement and gripper trajectory encountered in real experiments. As discussed in Section~\ref{sec:related-work}, the SOTA algorithms for evaluating the quality of a grasp configuration in active grippers use machine learning approaches, but these results cannot be directly applied for passive gripper design. The results in this work invites new avenues of research in this direction. Future work should also consider modeling external forces that may occur during pickup, for example when picking up an object from a bin.

\begin{figure}[tb]
    \centering
    \includegraphics[width=\columnwidth]{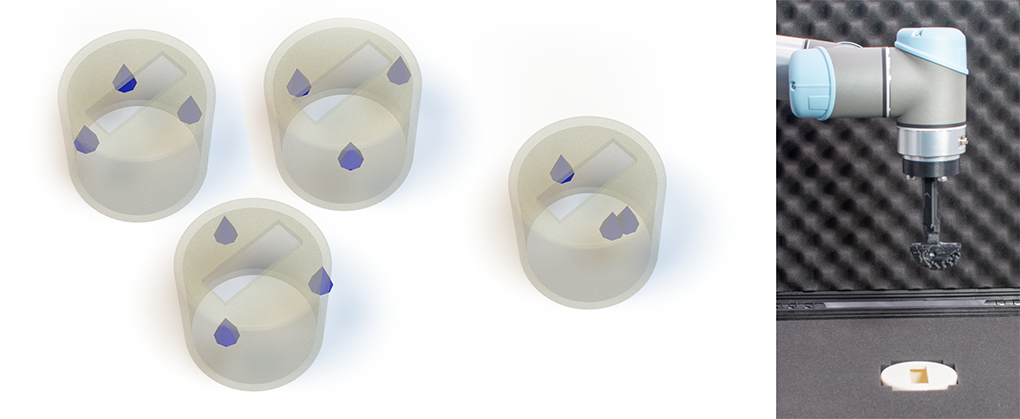}
    \caption{
    While our algorithm is able to grasp A3 with wrench and twist motion (see Fig.~\ref{fig:trajectory}), it cannot find a solution with insert and twist motion. In this experiment, we restricted the contact point to lie inside the object, and the only GCs found (see examples on the left) have no feasible solution since the algorithm prioritizes stability without knowing the shape of the slot. Given a manually specified GC (middle), our algorithm is able to find the solution (see example that has been validated in reality with 100\% reliability on the right). This example illustrates avenues of future work on GC and trajectory co-design.}
    \label{fig:topkey-cp}
\end{figure}

On the algorithmic side, while the random sampling of GC candidates shows numerous successes, they may be sub-optimal in terms of robustness. Future work should employ a better sampling algorithm for maximal robustness and diversity. Further analysis on the collision avoidance objective should also be done to find the optimal trade-off between computational time and convergence as we vary the significance of each cost component and reduce the sampling resolution. In particular, we observed that dropping the trajectory collision energy and the wrap-around distance achieves better convergence when the trajectory samples are sufficiently high. However, adding these components to the cost function is important for convergence with sparse samples.

It would be interesting to investigate alternatives to jointly optimize the contact point selection and trajectory optimization which cannot be completely decoupled (see Fig. \ref{fig:topkey-cp}). One other challenge is that the trajectory optimization can get stuck in a local minimum and not find a collision-free trajectory for a GC even if one exists. When this happens, the algorithm will choose a lower ranked GC. Additional freedom in object placement could also be achieved by optimizing robot's state in the grasping position.

Future work should also investigate different post-grasp trajectories. By assuming a vertical motion, we are unable to grasp objects with no bottom support such as a cylinder or a cone. Such objects can be handled by incorporating complex grasping strategies, such as rotating the object after being grasped~\cite{mucchiani2018object,mucchiani2021dynamic}. 

Finally, while we argue that customization for a single input has important practical applications, it would be interesting to relax some of the assumptions on our input. For example, we improve grasp reliability by considering random deviations of the object's position and geometry. Another interest direction for future work is to build on our optimization techniques to handle two or more input shapes or classes of shapes. 

%% file: sec/10_conclusion.tex
\section{Conclusion}

This work introduces a novel application for a generative design that has a high potential for impact in industrial applications of robotics. Robotic systems in industrial settings are highly inflexible---any change requires expensive re-design of autonomous components. The true cost of this rigidity has been particularly apparent during recent changes in demand during the COVID-19 crisis. Our algorithm addresses these challenges by establishing a new framework for creating robotic systems that can easily adapt to different scenarios without requiring additional dexterity, programming, or system complexity. Our key insight is that we can achieve this flexibility through re-design of passive grippers. We address the challenges of exploring the complex design space of passive grippers with novel insights on representation abstractions and co-design. We validate our findings with extensive physical experiments and discuss limitations, paving the way to exciting avenues of future work in this domain.